\documentstyle[prd,aps,preprint]{revtex}
\tightenlines
\input epsf.tex
\def\DESepsf(#1 width #2){\epsfxsize=#2 \epsfbox{#1}}

\begin{document}

\draft
\preprint{\vbox{
\hbox{OSU-HEP-99-06}
\hbox{UMD-PP-99-116} }}
\title{Supersymmetry, Local Horizontal Unification, \\ and a Solution
to the Flavor Puzzle}

\author{ K. S. Babu$^1$ and R. N. Mohapatra$^2$ }

\address{$^1$Department of Physics, Oklahoma State University,
Stillwater, OK, 74078, USA \\
$^2$Department of
Physics, University of Maryland, College Park, MD, 20742, USA}
\maketitle
\begin{abstract}
Supersymmetric gauge models with local horizontal symmetries are
known to generate large flavor changing neutral current
effects induced by supersymmetry breaking $D$--terms. We show how the
presence of a $U(1)$ gauge symmetry solves this problem.
We then construct a realistic gauge model  with $SU(2)_H\times U(1)_H$
as the local horizontal symmetry and suggest that the $U(1)_H$ factor
may be identified with the anomalous $U(1)$ induced by string compactification.
This model explains the observed hierarchies among the quark masses and
mixing angles, accommodates naturally the solar and atmospheric neutrino data,
and provides simultaneously a solution to the supersymmetric flavor problem.
The model can be excluded if the rare decay $\mu \rightarrow e \gamma$
is not observed in the current round of experiments.
\end{abstract}
\vskip0.5in

\section{Introduction}

One of the fundamental puzzles of the standard model is a lack of
understanding of the fermion mass and mixing hierarchies.
A promising approach to resolve
this puzzle is to use horizontal symmetries,
either global and local, that transform fermions of one generation
into another\cite{all}. The fact that
in the limit of vanishing Yukawa couplings, the standard model has an
enormous $[SU(3)_L^q\times SU(3)_R^u \times SU(3)_R^d \times SU(3)_L^\ell \times
SU(3)_R^e \times U(1)_{B-L}]$ horizontal symmetry makes this approach
quite plausible.

In this paper we shall be concerned with $local$ realization of
horizontal symmetries.  Local symmetries have certain clear--cut
advantages over their global counterparts.  The most significant
difference is perhaps the strong suspicion that global (but not local)
symmetries are susceptible to explicit violation through
quantum gravitational effects.

The presence of supersymmetry appears to put additional constraints on
models with $local$ horizontal symmetries\cite{mura}. For instance, if
one chooses a general simple non--Abelian group $G$ as the local horizontal
symmetry, in the presence of
supersymmetry breaking, the $D$--terms associated with $G$ will
induce a non--negligible splitting
among the slepton and squark masses of different generations. In turn
they induce unacceptable
amount of flavor changing neutral current effects. Since the squark mass
splittings are independent of the horizontal gauge couplings as long as the
horizontal group is simple, there is no free
parameter that can be used to dial down these splittings to an acceptable
level. This poses a serious roadblock to the use of local horizontal
symmetries as a way to understand the fermion mass hierarchies in
 a supersymmetric context.

If the $D$--term mass splittings between squarks (or sleptons)
were absent (or if they were under control), local horizontal
symmetries can neatly address the SUSY flavor problem
that plagues the generic softly broken supersymmetric standard
model.  Such a solution would be highly desirable
especially in scenarios where supersymmetry breaking is
communicated to the MSSM sector through supergravity.  The clear
advantage is that no special assumption need be made
about the Kahler potential (or the superpotential),
apart from the requirement of gauge invariance. Two different
approaches have been adopted in the past that evade the aforementioned
$D$--term difficulty: (i) assume the horizontal symmetry to be
$global$ \cite{dine}, or
(ii) use a discrete gauge symmetry \cite{kaplan}. In both cases there are no
associated $D$--terms.

In this paper we propose a solution to the $D$--term problem
that would facilitate the use of {\it true  gauge symmetries}
 to address simultaneously the fermion mass problem and the supersymmetric
flavor problem\footnote{For a specific
gauged $SO(3)$ model with symmetry breaking via fundamental
fields, see Ref. \cite{bb}.}.
We  show that by adjoining a local $U(1)$
symmetry to the existing non--Abelian horizontal symmetry $G$, the mass
splittings between different generations of squarks caused by the $D$--terms
can be brought under control. The mass splittings will now depend quadratically
on the ratio of the two gauge couplings, which
can be adjusted to make the FCNC effects sufficiently small.
We illustrate this mechanism by using the example of a horizontal
$SU(2)_H\times U(1)_H$ model. We then construct a fully realistic
model using this group and show how the $U(1)$ factor may be
identified with the anomalous $U(1)$ that arises in superstring
compactification. We find this model to be quite predictive in the
neutrino as well as in the quark and lepton sectors.  Specifically,
it supports the large angle atmospheric and small
 angle solar neutrino oscillations.
The model can be excluded if the rare decay $\mu \rightarrow e \gamma$ is not
observed in the current round of experiments.
Furthermore, this model can easily be grand unified into an $SU(5)$ or
$SO(10)$ group.

\section{Suppression of $D$--term splittings in the presence of an extra $U(1)$}

Let us consider the gauge group of the theory to be $SU(3)_C\times SU(2)_L
\times U(1)_Y\times G_H$, where the horizontal group $G_H$ is chosen to be
$SU(2)_H\times U(1)_H$ as stated before. We will choose the matter content
of the model to be the
same as that of the MSSM with the addition of three
right--handed neutrinos (denoted by $\nu^c_i$).  These $\nu^c_i$ fields are
required for the cancellation of triangle and global $SU(2)_H$ anomalies.
An immediate consequence is that the left--handed neutrinos will have
small masses induced by the seesaw mechanism.
The horizontal quantum
numbers are chosen under the straightforward assumption that the
particles of the first two generation belong to doublets of the
$SU(2)_H$ group whereas the particles of the third generation are in the
singlet representation. (This assignment is same as in Ref.
\cite{dine}).  Furthermore, we impose the unifiability condition, which
means that all 16 members of a generation have the same horizontal charges.
Thus the matter content of the model in the standard notation is:
\begin{eqnarray}
\{Q_a, ~L_a,~ u^c_a,~ d^c_a,~ e^c_a,~ \nu^c_a \} &:& {\bf 2}(-1)
\nonumber \\
\{Q_3, ~L_3,~ u^c_3,~ d^c_3,~ e^c_3,~ \nu^3_a \} &:& {\bf 1}(0)~
\end{eqnarray}
where we have exhibited the $SU(2)_H \times U(1)_H$ quantum numbers.  Here
$a = 1-2$ is the $SU(2)_H$ index.  The Higgs sector consists of the following
fields:
\begin{eqnarray}
\{H_u,~H_d\} : {\bf 1}(0); ~~ \phi_a: {\bf 2}(1);~~ \overline{\phi}_a:
{\bf 2}(-1);~~
\chi: {\bf 1}(1); ~~\overline{\chi}: {\bf 1}(-1);~~S_i: {\bf 1}(0) ~(i=1-2)~.
\end{eqnarray}
Here $H_u,H_d$ are the usual MSSM doublet fields, while all the other fields are
singlets of the standard model.  Given the horizontal symmetry
group $G$, this Higgs system is the minimal choice that can properly break
$G$ without breaking supersymmetry.  The $(\phi,\overline{\phi})$ fields break
$SU(2)_H$ completely, while the $(\chi, \overline{\chi})$ fields break $U(1)_H$.
The $S_i$ fields are necessary to allow cubic terms in
the superpotential, a requirement if the horizontal gauge symmetry is to
be broken in the supersymmetric limit with only renormalizble terms.

Let us first demonstrate how the suppression of the $D$--term mass splittings
between different generation squarks (and sleptons) arises in this model.
 In this
particular model the problem concerns the $SU(2)_H$ $D$--terms which can
potentially split the masses of the first two generations.
Note that the $U(1)_H$ $D$--term will not induce mass
splittings within the first two generations.  The most general
superpotential involving the Higgs fields has the form:
\begin{equation}
W = \mu_\phi \phi \overline{\phi} + \lambda \phi \overline{\phi} S_1 +
W'(\chi \overline{\chi}, S_i)~,
\end{equation}
where we wont need the explicit form of the
piece $W'$ involving $(\chi \overline{\chi}, S_i)$ fields.  In the
supersymmetric limit, we have $\left\langle \phi \right \rangle
= \left \langle \overline{\phi} \right \rangle \equiv V_\phi$ and $\left \langle
\chi \right \rangle = \left \langle \overline{\chi} \right \rangle \equiv
 V_\chi$.
We shall make the reasonable assumption that the scale of horizontal
symmetry breaking $(V_\phi, V_\chi)$ is much greater than
the scale of soft supersymmetry
breaking terms, $M_{\rm SUSY}$.
The requirement of vanishing $F$--terms in the supersymmetric limit  implies
$F_\phi = (\mu_\phi + \lambda S_1)\overline{\phi} = 0$ and $F_S = \lambda \phi
\overline{\phi} + \partial W'/\partial S_1 = 0$.
  Including arbitrary soft supersymmetry
breaking, the scalar potential involving
 $(\phi, \overline{\phi})$ fields is given by:
\begin{eqnarray}
V &=& \left|\mu_\phi + \lambda S\right|^2(|\phi|^2+|\overline{\phi}|^2) +
\left|\lambda \phi \overline{\phi} + \partial W'/\partial S\right|^2 + {1
\over 8} (g_{2H}^2+4 g_{1H}^2)(|\phi|^2-|\overline{\phi}|^2)^2 \nonumber \\
&+& m_\phi^2 |\phi|^2 + m_{\overline{\phi}}^2|\overline{ \phi}|^2 +
\{B_\phi \mu_\phi \phi \overline{\phi} + A_\phi \lambda \phi \overline{\phi}S
+ H.c.\}~.
\end{eqnarray}
The coefficients in the last line of Eq. (4)
($m_\phi^2, m_{\overline{\phi}}^2, B_\phi^2,
A_\phi^2)$ are all of order $M_{\rm SUSY}^2$.
Here we have adopted the point of view that in the absence of a symmetry,
the superpartner masses at the Planck scale should be arbitrary.  Specifically,
we do not assume universality of scalar masses.

Minimizing this potential (Eq. (4))
with respect of $\phi$ and $\overline{\phi}$ fields and subtracting the two
extremization conditions,
we arrive at the relation:
\begin{equation}
(|\overline{\phi}|^2-|\phi|^2)\left[{\lambda^* \over \phi \overline{\phi}}
(\lambda \phi \overline{\phi} + \partial W'/\partial S)
-{1 \over 4} (g_{2H}^2+4g_{1H}^2) + B_\phi \mu_\phi + A_\phi \lambda S\right]
= (m^2_{\overline{\phi}}- m^2_\phi)~.
\end{equation}
Noting that $\lambda \phi \overline{\phi} + \partial W'/\partial S_1
 = {\cal O}(M_{\rm SUSY} V_\phi)$, Eq. (5) implies:
\begin{equation}
{ 1 \over 4} (g_{2H}^2+4 g_{1H}^2)(|\phi|^2 - |\overline{\phi}|^2) \simeq
-(m_\phi^2-
m^2_{\overline{\phi}}) + {\cal O}(M_{\rm SUSY}/V_\phi)M_{\rm SUSY}^2~.
\end{equation}

The contribution to the squark (or slepton) mass splittings from the $D$--term
is given by:
\begin{equation}
\Delta m^2_{\tilde{q}} \simeq {1 \over
4}g_{2H}^2(|\phi|^2-|\overline{\phi}|^2) \simeq
{g_{2H}^2 \over g_{2H}^2+4g_{1H}^2} (m^2_{\overline{\phi}}-m^2_\phi)~.
\end{equation}
Note that in the absence of the $U(1)_H$ group (i.e., if $g_{1H}=0$),
$\Delta m^2_{\tilde{q}}\simeq m^2_{\overline{\phi}}-m^2_\phi$ which is
independent of the gauge
coupling and is arbitrary (i.e., anywhere between $(100~{\rm
GeV})^2-(1000 ~{\rm GeV})^2$.
Although these $D$--terms contribute to diagonal squark masses, because
they are not universal, once Cabibbo rotation on the quark fields are made, they
do contribute to flavor changing processes.
The most stringent constraints arise from $K^0-\bar{K^0}$ mixing and the rare
decay $\mu \rightarrow e \gamma$.  The $K_L-K_S$ mass difference
sets a constraint\cite{masiero} (from the most stringent $(LL)(RR)$ operator)
$[{\Delta m^2_{\tilde{q}}/m^2_{\tilde{q}}}]
\theta_C \leq 1 \times 10^{-3}({m_{\tilde{q}}}/{500~ GeV})$, where
$m_{\tilde{q}}$
denotes the average squark mass.  The constraint arising from $\mu \rightarrow
e\gamma$ is similar.
Clearly, if $g_{1H} \rightarrow 0$,
the $D$--term splittings will grossly contradict
experiments\footnote{For an exception where $\theta_C$ arises almost
entirely from the up--quark sector, see Ref. \cite{nir}.} if the
squark masses are below a TeV.
On the other hand, in the presence of the extra gauge coupling
$g_{1H}$, we can control the FCNC processes to adequate levels.  For example,
if $g_{2H}/g_{1H} = (1/3-1/7)$, which is not at all unreasonable, then
$\Delta m^2_{\tilde{q}} \simeq (1/40-1/200)(m^2_{\tilde{\phi}}-m^2_\phi)$.  For
$m_{\tilde{q}} \sim (300-500)$ GeV, this is comfortably consistent with
experiments.
Note that the soft supersymmetry breaking mass terms $m^2_\phi$ and
$m^2_{\overline{\phi}}$
do not run below the horizontal scale (since the $\phi$ and
$\overline{\phi}$ fields
have masses of order $V_\phi$).  On the other hand, the masses of the squarks do
run below $V_\phi$ and  in this process receive
a flavor universal contribution from the gauginos.  For comparable values
of initial (Planck scale) gaugino and squark masses, the factor
$m^2_\phi/m^2_{\tilde{q}}$ is
suppressed by about $\sim 1/10$ because of the running.

While we used a specific example to illustrate the proposed mechanism to
cure the
$D$--term problem, its features prevail in more general contexts.  For example,
we could use alternative superpotentials (Cf. Eq. (3)) such as $W =
\lambda S_1(\phi
\overline{\phi}-\mu^2) + W'$ or one involving non--renormalizable
operators.  The FCNC
problem in the absence of $U(1)_H$, and its solution via $U(1)_H$
will be identical to the case discussed above.

\section{A realistic model of fermion mass and mixing hierarchies}

We will now show that the model presented in the previous section can naturally
explain the fermion mass and mixing hierarchies.  We take the viewpoint that
all operators consistent with gauge invariance are allowed in the Lagrangian.
This includes non--renormalizable operators, which will be suppressed by
appropriate inverse powers of the Planck mass.  The coefficients of such
operators will all be assumed to be of order unity.

Let us first consider the quark sector of the theory.
It is easy to see that the superpotential consistent with
$SU(2)_H \times U(1)_H$ symmetry is given by:
\begin{eqnarray}
W_{\rm Yuk} &=& h_{33}^u Q_3u^c_3H_u + h_{33}^d Q_3 d^c_3 H_d +\frac{
h_{23}^u}{M}
\epsilon^{ab} Q_au^c_3 H_u \phi_b + {h_{23}^d \over M} \epsilon^{ab} Q_a
d_3^c H_d \phi_b
+ {h_{32}^u \over M} \epsilon^{ab} Q_3 u^c_a H_u \phi_b \nonumber \\
&+& {h_{32}^d \over M}
\epsilon^{ab} Q_3 d_a^c H_d \phi_b + {h_{22}^u \over M^2}Q_a u_b^c H_u
\phi_p\phi_q
\epsilon^{ap} \epsilon^{bq} + {h_{22}^d \over M^2}Q_a d_b^c H_d \phi_p \phi_q
\epsilon^{ap}\epsilon^{bq} \nonumber \\
&+& {h_{12}^u \over M^2} \epsilon^{ab}Q_a u_b^c H_u \chi^2
+ {h_{12}^d \over M^2} \epsilon^{ab} Q_a d_b^c H_d \chi^2~.
\end{eqnarray}
Defining two small parameters $\epsilon_{\phi}\equiv \frac{\left \langle\phi
\right \rangle}{M}$
and $\epsilon{\chi}\equiv \frac{\left\langle \chi \right \rangle}{M}$, we get the following
hierarchical mass matrix for both up and the down sectors:
\begin{eqnarray}
M_f~=~v_f\left(\begin{array}{ccc}
0  & h_{12}^f\epsilon^2_{\chi} & 0 \\
-h_{12}^f \epsilon^2_{\chi} & h_{22}^f\epsilon^2_{\phi} & h_{23}^f\epsilon_{\phi} \\
0 & h_{32}^f\epsilon_{\phi} & h_{33}^f \end{array}\right)~
\end{eqnarray}
with $f=u,d$.  The charge lepton mass matrix has an identical form as Eq.
(9), as does the Dirac neutrino mass matrix (identify $f=\ell, \nu$ for
the two cases).

The mass matrices in Eq. (9) naturally explain the fermion mass and mixing angle
hierarchies.  To see this in detail, assume that all the $h_{ij}$
parameters are of order one.
The mass ratios in the down--quark sector is then given by:
\begin{equation}
m_s/m_b \sim \epsilon_\phi^2,~~~ m_d/m_s \sim \epsilon_\chi^4/\epsilon_\phi^4
\end{equation}
with similar results in the up--quark and the charged lepton sectors.  If
we choose
$\epsilon_\phi \simeq 1/7$ and $\epsilon_\chi \simeq 1/20$, all the
observed masses
and mixing angles can be explained, with the coefficients $h_{ij}$
taking values in the range $(1/2 -2)$.  This is a tremendous improvement
over the
standard model Yukawa couplings which span some six orders of magnitude.
In our scheme, order one differences  such as in
$m_\mu/m_\tau$ and $m_s/m_b$ (the two differ by about a factor of 3 near
the Planck scale)
are attributed to order one differences in the $h_{ij}$ couplings.  The
hierarchy
$m_b/m_t$ requires moderate to large values of $\tan\beta\sim 10-40$.

The zeros in the mass matrices of Eq. (9) are corrected only at very high order,
and are negligible.
(The (1,1) entry receives a correction of order $\epsilon_\phi^2
\epsilon_\chi^4$,
the (1,3) and (3,1) entries are corrected at order $\epsilon_\phi
\epsilon_\chi^2$.)
The near vanishing of the (1,1) entry, along with the (anti)-symmetry of the
(1,2) entry leads to a successful quantitative prediction for the Cabibbo angle;
the vanishing of the (1,3) and (3,1) entries yield a relation for
$V_{ub}$ (and $V_{td}$):
\begin{equation}
|V_{us}| \simeq \left|\sqrt{m_d/m_s}- \sqrt{m_u/m_c}e^{i \alpha}\right|~;~~~
|V_{ub}|/|V_{cb}| \simeq \sqrt{m_u/m_c}~;~~~ |V_{td}|/|V_{ts}| \simeq \sqrt{m_d/m_s}~.
\end{equation}
The last two relations \cite{rasin} could serve as future tests of the model.

Turning now to the leptonic sector, as noted, the charged lepton and the
Dirac neutrino
mass matrices have identical forms as Eq. (9).  The $\nu^c_i$ Majorana
mass matrix is
obtained from the Lagrangian:
\begin{eqnarray}
{\cal L}^{\nu^c} &=& f_{33}\nu_3^c \nu_3^c \Delta +{f_{23} \over M}
\epsilon^{ab} \nu_a^c \nu_3^c \Delta \phi_b + {f_{22} \over M^2} \nu^c_a \nu^c_b
\Delta \phi_p \phi_q
\epsilon^{ap} \epsilon^{bq} \nonumber \\
&+& {f_{13} \over M^3}
\epsilon^{ab}\nu_a^c \nu_3^c \Delta \overline{\phi}_b \chi^2 +
{f_{12} \over M^4} \nu_a^c \nu_b^c \Delta \overline{\phi}_p \overline{\phi}_q
\epsilon^{ap} \epsilon^{bq} \chi^2 + {f_{11} \over M^6} \nu_a^c \nu_b^c \phi_p
\overline{\phi}_q \epsilon^{ap}\epsilon^{bq} \chi^4~.
\end{eqnarray}
At the level of the standard model,
a bare mass term will be allowed for the $\nu_3^c$ Majorana mass.  We have
assumed it to arise from the VEV of a field $\Delta$ that breaks $B-L$ symmetry.
When the model is embedded into a left--right symmetric or $SO(10)$ framework,
or if the $B-L$ symmetry of the model as it stands is gauged, the Majorana
masses of the $\nu^c$ fields will require the VEV of such a multiplet.  Apart
from the motivation to unify, we follow this path since then $R$--parity
violating terms will be automatically eliminated from the Lagrangian.

Due to the intricacy of the  seesaw diagonalization, we
have kept the lowest non--vanishing terms in all entries of the Majorana
$\nu^c$ matrix, which is given by:
\begin{eqnarray}
M_{\nu^c}~=~\left\langle \Delta \right\rangle\left(\begin{array}{ccc}
f_{11}\epsilon^2_{\phi}\epsilon^4_{\chi} & f_{12} \epsilon_\phi^2
\epsilon_\chi^2 &  f_{13}\epsilon_{\phi}\epsilon^2_{\chi}
\\
f_{12}\epsilon_\phi^2\epsilon_\chi^2 & f_{22}\epsilon^2_{\phi} &
f_{23}\epsilon_{\phi} \\
f_{13}\epsilon_{\phi}\epsilon^2_{\chi} & f_{23}\epsilon_{\phi} & f_{33}
\end{array}\right)~.
\end{eqnarray}
Using the seesaw formula the light left--handed neutrino mass matrix is obtained to be:
\begin{eqnarray}
M^{\rm light}_{\nu}~=~-\frac{v_u^2}{\left\langle \Delta \right \rangle}
\left(\begin{array}{ccc}
F_{11}\epsilon_\chi^4/\epsilon_\phi^{2} & F_{12}\epsilon_\chi^2/\epsilon_\phi^{2} &
F_{13}\epsilon_\chi^2/\epsilon_\phi \\
F_{12} \epsilon_\chi^2/\epsilon_\phi^{2} & F_{22} \epsilon_\phi^{-2} & F_{23}
\epsilon_\phi^{-1} \\
F_{13} \epsilon_\chi^2/\epsilon_\phi & F_{23}\epsilon_\phi^{-1} &  F_{33}
\end{array}\right)~.
\end{eqnarray}
Here $F_{ij}$ are functions of various combinations of $h_{ij}^\nu$ and $f_{ij}$,
and are expected to be of order one.

It is amusing to note that the largest entry in Eq. (14) corresponds to
the mass of $\nu_\mu$.
The light neutrino mass hierarchy predicted in the model is $m_{\nu_e} \ll
m_{\nu_\tau} \ll m_{\nu_\mu}$.
If we set $\left\langle \Delta \right \rangle = 2 \times 10^{16}$ GeV, which is
the supersymmetric unification scale, $m_{\nu_\mu} \approx 7 \times 10^{-2}$
eV.  This is in the right range to explain the atmospheric neutrino data via
$\nu_\mu-\nu_\tau$ oscillations.  The relevant mixing angle is given by
$\theta_{\mu\tau}^{\rm osc} \simeq (F_{23}/F_{22} - h_{23}^\ell/h_{33}^\ell)\epsilon_\phi.$
This angle can be near maximal ($45^0$) if, for example, $F_{23}/F_{22}\sim 3-4$.
This appears quite plausible, given that $F_{ij}$ are non--trivial combinations
of the Yukawa couplings.  $\{F_{23}/F_{22} = (f_{13}f_{22}h_{33}^\nu
+ f_{12}f_{23} h_{33}^\nu - f_{12}f_{33}h_{32}^\nu+
f_{13}f_{23}h_{32}^\nu)/[(h_{12}^\nu(f_{23}^2-f_{22}f_{33})]\}$.
As for the solar neutrino problem, it is explained by small angle
$\nu_e-\nu_\tau$ MSW oscillations.  The mass of $\nu_\tau$ is of order
$\epsilon_\phi^2 \times m_{\nu_\mu} \sim 10^{-3}$ eV.  The $\nu_e-\nu_\tau$
mixing angle is of order $\epsilon_\chi^2/\epsilon_\phi \simeq 0.02$.  Both
parameters neatly fit the desired values \cite{bilenkii}.   Large angle
$\nu_e$ oscillations
are very unlikely in this model, if it is established the model could be
excluded.

\section{Solving the supersymmetric flavor problem}

The model presented here has a built--in solution to the supersymmetric
flavor problem.
In fact, part of our motivation to use the horizontal symmetry was to address
this problem.  Since the horizontal symmetry is local, no explicit violation
is expected from quantum gravitational effects.
As already discussed (see Sec. II),
augmenting the horizontal symmetry group by a $U(1)$ factor alleviates flavor
violation arising from the horizontal $SU(2)_H$ $D$--terms.
Flavor violation in the squark (and the slepton) sector will however be
induced, once effects of horizontal symmetry breaking are included.  We
will now show that such violations are not excessive and are consistent
with present FCNC constraints.

We assume that supersymmetry breaking occurs in a hidden sector and
is communicated to the MSSM sector by supergravity.  However, we do not
make any special assumption about the Kahler potential or the superpotential.
In particular, we do not assume that the scalar masses are universal or that
the supersymmetry breaking
trilinear $A$--terms are proportional to the superpotential Yukawa couplings.
The soft scalar masses  arise from  general Kahler potential terms.
For the first two generation squarks the dominant contribution arises from
\begin{equation}
{\cal L} = \int d^4 \theta Q_a^\dagger Q_a {z^* z \over M_{\rm Planck}^2}~
\end{equation}
where $z$ is a hidden sector (spurion) field with nonzero $F$--component that
breaks supersymmetry.  Identifying $F_z^2/M_{\rm Planck}^2
 \equiv M_{\rm SUSY}^2$, we
see that the dominant masses for the first two generations are universal as
a consequence
of the $SU(2)_H$ horizontal symmetry.  Non--universal corrections will
arise from terms such as
\begin{equation}
{\cal L} = \int d^4 \theta (Q_a^\dagger \phi_a) (\phi_b^\dagger Q_b)
{z^* z \over M_{\rm Planck}^4}~
\end{equation}
and a similar term with $\phi$ is replaced by $\overline{\phi}$.
Compared to the dominant contribution (Eq. (15), these non--universal terms are
suppressed by a factor $\epsilon_\phi^2 \sim 2 \times 10^{-2}$.  Since
these corrections contribute to diagonal entries in the squark mass matrix, any
FCNC effect will require an additional quark mixing angle ($\sim \theta_C
\sim \sqrt{m_d/m_s} \simeq 0.2$).  Furthermore, as noted earlier,
these non--universal corrections get diluted by about a factor of 1/10 in
the squark sector through RGE effects proportional to the gaugino (mainly the
gluino) mass.  It is convenient to  define a set of parameters
$\delta_{12}^d$ as the ratio of the (1,2) entry of the
squark mass matrix to the average squark mass--squared in a basis where
the quark fields are physical \cite{masiero}.  We then
estimate $(\delta_{12}^d)_{LL,RR} \sim (2 \times 10^{-2})
\times (0.2) \times (0.1) = 4 \times 10^{-4}$.  This is to be compared with the
experimental limit on this quantity, $(\delta_{12}^d)_{LL,RR} \le 1
\times 10^{-3}$
valid for an average squark mass of 500 GeV \cite{masiero}.  We see broad
agreement with experiment.

Analogous discussion in the first two generation slepton sector leads to
a prediction
$(\delta_{12}^\ell)_{LL,RR} \simeq \epsilon_\phi^2 \theta_{e\mu}
\simeq (2 \times 10^{-2}) \times (0.07)
\simeq 1.4 \times
10^{-3}$.  Here we have taken the $e-\mu$ mixing angle to be
$\sqrt{m_e/m_\mu} \simeq 0.07$,
appropriate to the mass matrix of Eq. (9).  Note that unlike the squark
sector, there
is no significant dilution effect due to the RGE (since sleptons are
color neutral).
This number should be compared with the constraint from the present experimental
limit on $\mu \rightarrow e \gamma$, which is $(\delta_{12}^\ell)_{LL,RR} \le
(4.0 \times 10^{-3} - 1.8 \times 10^{-2})$ corresponding to $m_{\tilde{\ell}}
\simeq $ 100 GeV and for $x \equiv m_{\tilde{\gamma}}^2/m_{\tilde{\ell}}^2 $
in the range  $0.3-3$ \cite{kim}.  Although the constraint is satisfied,
the rate for $\mu \rightarrow e \gamma$ cannot be much below the
present experimental limit.  We estimate the rate to be at most a factor of
100 below the present limit, which will soon be tested.

As for the supersymmetry breaking trilinear $A$ terms, they arise in
supergravity models from superpotential terms such as
\begin{equation}
{\cal L} = \int d^2\theta Q_3 d_3^c H_d {z \over M_{\rm Planck}}~.
\end{equation}
The resulting coefficients of the trilinear scalar terms are of order
$M_{\rm SUSY}$. Horizontal gauge invariance implies that in our model, the
structure of the $A$--terms is identical to that of the superpotential in
Eq. (8).  However, the coefficients of the $A$ matrix are not proportional
to the Yukawa matrix arising from Eq. (8). This non--proportionality will
lead to FCNC processes.  We estimate the parameter
$(\delta_{12}^d)_{LR} \sim \epsilon_\chi^2 A v_d/m_{\tilde{q}}^2$.  Choosing
$m_{\tilde{q}} = 500$ GeV and $\tan\beta = 30$, we obtain $(\delta_{12}^d)_{LR}
\simeq 3 \times 10^{-5}$, which is well below the experimental limit on this
quantity arising from $K^0-\bar{K^0}$ mixing
$((\delta_{12}^d)_{LR} \le 3 \times 10^{-3}$).  A similar estimate
(perhaps slightly smaller value, since $\sqrt{m_e/m_\mu} \simeq (1/3)\sqrt{m_d/m_s}$)
will hold for the leptonic $(\delta_{12}^\ell)_{LR,RL}$.  For slepton masses of
500 GeV, the constraint from $\mu \rightarrow e \gamma$ is
$(\delta_{12}^\ell)_{LR,RL} \le 2 \times 10^{-5}$.  We see that the constraint
is quite tight.  Allowing for unknown order one coefficients (or some
proportionality of the $A$ terms) we conclude that $\mu \rightarrow e \gamma$ cannot be
much below
the present experimental limit.  Since the coefficients of $\tilde{\mu}_L
\tilde{e}_R$
and $\tilde{\mu}_R \tilde{e}_L$ are approximately the same, we expect that both
helicity muons will participate in the decay, unlike the grand unification
effects discussed in Ref. \cite{barbieri} where the decay $\mu_L \rightarrow e_R
\gamma$
is suppressed.  The rate for the decay $\tau \rightarrow \mu \gamma$ is estimated
to be two orders of magnitude below the present limits.

\section{Comments and conclusions}

Before concluding a few remarks are in order.

(i)  An important check for the renormalizability of the model is that the
horizontal gauge symmetries be anomaly free.  $SU(2)_H$ is automatically
free of chiral anomalies.  Cancellations of anomalies involving the $U(1)_H$
should however be ensured.  With the assignment of $U(1)_H$ charges as given, we
find the anomaly coefficients for $U(1)_H \times [SU(2)_L]^2$, ~$U(1)_H
\times [SU(3)_C]^2$,
$U(1)_H\times [SU(2)_H]^2$ and $U(1)_H \times [U(1)_Y]^2$ are all equal to $-8$.
 (We have used the GUT normalization for the $Y$ quantum number.)  It is
then very
tempting to identify the $U(1)_H$ as the pseudo--anomalous $U(1)$ that arises
in superstring compactification \cite{witten}.
An attractive aspect of this identification is that the $\chi$ and
$\bar{\chi}$ fields can play the role of the singlet fields for the
purpose of model building\cite{dvali} and the dominant mode of
supersymmetry breaking could be via the $D$--terms of the anomalous $U(1)$
group. In that case, one of them (which can be chosen to be the $\chi$
field) picks up a VEV of order
$\frac{1}{10}M_{\rm Planck}$. Thus if we scale all higher dimensional operators
by the Planck mass, we get the desired order for the $\epsilon_{\chi}$.
It is then clear that we must choose $\left\langle \phi\right\rangle\simeq
\left\langle\bar{\phi}\right\rangle$ also of
the same order.  This identification will go well with the fits obtained
from the fermion masses.
The $\bar{\chi}$ could be used to break supersymmetry
if we kept only the $\chi\bar{\chi}$ term in the superpotential as in
Ref. \cite{dvali}. We do not pursue this line here.  It might
be mentioned that anomaly cancellation can also occur by introducing
fields which are vector--like under the standard model, but chiral under
the $U(1)_H$.

(ii)  The model as it stands can easily be embedded into the grand unification
groups such as $SU(5)$ or $SO(10)$ since the horizontal quantum numbers
are common to all the quarks and leptons that fit into a single multiplet
of the above groups. As usual, for the case of $SU(5)$
unification the MSSM doublets must be embedded into {\bf 5} and $\bar{\bf
5}$ representations of $SU(5)$. For the case of $SO(10)$
grand unification, additional multiplets belonging to {\bf 126} or {\bf
16} will be needed.

(iii) CP violation in the model can arise via complex Yukawa couplings as in the
standard model.  There are additional supersymmetric source for CP violation.
An interesting possibility is that $\epsilon'/\epsilon$ in the $K$ meson system
can be explained through the gluino penguin.  As noted earlier, the parameter
$(\delta_{12}^d)_{LR,RL} \simeq 3 \times 10^{-5}$ in our model.  If its
argument is of order one, it is of the right magnitude to explain
$\epsilon'/\epsilon \simeq 2.7 \times 10^{-3}$ \cite{bdm}.
$(\delta_{11}^d)_{LR,RL}$,
which arises after Cabibbo rotation, will be of order $6 \times 10^{-6}$.
This will lead to a neutron (and electron) electric dipole moment very near the present
experimental limit.  It should, however, be noted that the horizontal symmetry
by itself does not fully cure the electric dipole moment problem, since
the $B$ parameter and the gaugino masses should have small phases.
An interesting scenario would be one where SUSY breaking terms arising
through the superpotential only (the $A$--terms) have order one phases.

In conclusion, we have discussed a way to avoid excessive FCNC effects
induced through $D$--term mass splittings between
squarks and sleptons  of different generations
in models with local horizontal symmetries.
We have constructed an anomaly free model with $SU(2)_H\times U(1)_H$
as the local horizontal symmetry group and shown that it can lead to
a proper understanding of the observed hierarchies among the quark and
lepton masses and their mixings.  We have shown that the model
provides a simultaneous
solution to the supersymmetric flavor problem.
Without any additional assumption this
model also leads to a desirable pattern of neutrino masses and mixings;
it supports small angle oscillations for the solar and large angle oscillations
for the atmospheric neutrino data.  Although flavor violation in the
model is under control, it is not unobservable.  The
rare decay $\mu \rightarrow
e \gamma$ is predicted to be near the present experimental limit.

The work of K.S.B is supported by funds from the Oklahoma State University.
R.N.M. is supported by the National Science Foundation grant No. PHY-9802551.

\end{document}